\documentclass{icrc2009}

\usepackage{graphicx}   
\usepackage[caption=false]{caption}    
\usepackage[font=footnotesize]{subfig} 
\usepackage{fixltx2e}
\usepackage{url}

\newcommand{\shorttitle}[1]%
{\markboth{Proceedings of the 31\MakeLowercase{$^{st}$} ICRC, {\L}\'{o}d\'{z} 2009}{#1} }
\newcommand{\etal}{\MakeLowercase{\textit{et al. }}} 

\begin{document}
\title{Search for a diffuse flux of high-energy
neutrinos with the Baikal neutrino telescope NT200}

\author{\IEEEauthorblockN{A. Avrorin\IEEEauthorrefmark{1},
			  V. Aynutdinov\IEEEauthorrefmark{1},
                          V. Balkanov\IEEEauthorrefmark{1},
                          I. Belolaptikov\IEEEauthorrefmark{4},
			  D. Bogorodsky\IEEEauthorrefmark{2},
                          N. Budnev\IEEEauthorrefmark{2},\\
                          I. Danilchenko\IEEEauthorrefmark{1},
                          G. Domogatsky\IEEEauthorrefmark{1},
			  A. Doroshenko\IEEEauthorrefmark{1},
                          A. Dyachok\IEEEauthorrefmark{2},
			  Zh.-A. Dzhilkibaev\IEEEauthorrefmark{1},\\
                          S. Fialkovsky\IEEEauthorrefmark{6},
			  O. Gaponenko\IEEEauthorrefmark{1},
                          K. Golubkov\IEEEauthorrefmark{4},
                          O. Gress\IEEEauthorrefmark{2},
			  T. Gress\IEEEauthorrefmark{2},
                          O. Grishin\IEEEauthorrefmark{2},\\
			  A. Klabukov\IEEEauthorrefmark{1},
                          A. Klimov\IEEEauthorrefmark{8},
                          A. Kochanov\IEEEauthorrefmark{2},
                          K. Konischev\IEEEauthorrefmark{4},
                          A. Koshechkin\IEEEauthorrefmark{1},
                          V. Kulepov\IEEEauthorrefmark{6},\\
                          D. Kuleshov\IEEEauthorrefmark{1},
                          L. Kuzmichev\IEEEauthorrefmark{3},
                          V. Lyashuk\IEEEauthorrefmark{1},
                          E. Middell\IEEEauthorrefmark{5},
                          S. Mikheyev\IEEEauthorrefmark{1},
                          M. Milenin\IEEEauthorrefmark{6},\\
                          R. Mirgazov\IEEEauthorrefmark{2},
                          E. Osipova\IEEEauthorrefmark{3},
                          G. Pan'kov\IEEEauthorrefmark{2},
                          L. Pan'kov\IEEEauthorrefmark{2},
                          A. Panfilov\IEEEauthorrefmark{1},
                          D. Petukhov\IEEEauthorrefmark{1},\\
                          E. Pliskovsky\IEEEauthorrefmark{4},
                          P. Pokhil\IEEEauthorrefmark{1},
                          V. Poleschuk\IEEEauthorrefmark{1},
                          E. Popova\IEEEauthorrefmark{3},
                          V. Prosin\IEEEauthorrefmark{3},
                          M. Rozanov\IEEEauthorrefmark{7},\\
                          V. Rubtzov\IEEEauthorrefmark{2},
                          A. Sheifler\IEEEauthorrefmark{1},
                          A. Shirokov\IEEEauthorrefmark{3},
                          B. Shoibonov\IEEEauthorrefmark{4},
			  Ch. Spiering\IEEEauthorrefmark{5},
			  O. Suvorova\IEEEauthorrefmark{1},\\
			  B. Tarashansky\IEEEauthorrefmark{2},
			  R. Wischnewski\IEEEauthorrefmark{5},
			  I. Yashin\IEEEauthorrefmark{3},
			  V. Zhukov\IEEEauthorrefmark{1}}
                            \\
\IEEEauthorblockA{\IEEEauthorrefmark{1}Institute for Nuclear Research of Russian Academy of Sciences,\\
     117312, Moscow, 60-th October Anniversary pr. 7a, Russia}
\IEEEauthorblockA{\IEEEauthorrefmark{2}Irkutsk State University, Russia}
\IEEEauthorblockA{\IEEEauthorrefmark{3}Skobeltsyn Instutute of Nuclear Physics MSU, Moscow, Russia}
\IEEEauthorblockA{\IEEEauthorrefmark{4}Joint Institute for Nuclear Research, Dubna, Russia}
\IEEEauthorblockA{\IEEEauthorrefmark{5}DESY, Zeuthen, Germany}
\IEEEauthorblockA{\IEEEauthorrefmark{6}Nizhni Novgorod State Technical University, Nizhnij Novgorod, Russia}
\IEEEauthorblockA{\IEEEauthorrefmark{7}St.Petersburg State Marine University, St.Petersburg, Russia}
\IEEEauthorblockA{\IEEEauthorrefmark{8}Kurchatov Institute, Moscow, Russia}}

\shorttitle{A. Avrorin \etal Search for a diffuse flux of high-energy
neutrinos}
\maketitle

\begin{abstract}
We present the results of a new analysis of data taken in 1998-2002 for
a  search for high-energy  extraterrestrial
neutrinos. The analysis is based on a full reconstruction of high-energy
cascade parameters: vertex coordinates, energy and arrival direction. 
Upper limits on the diffuse fluxes of all neutrino flavors,
predicted by several models of AGN-like neutrino sources are derived.
For an ${\bf E^{-2}}$ behavior of the neutrino spectrum, our limit is
${\bf E^{2} F_{\nu}(E) < 2.9 \times 10^{-7}}$ cm${\bf ^{-2}}$ s${\bf ^{-1}}$ sr${\bf ^{-1}}$
GeV over a neutrino energy range ${\bf 2 \times 10^4 \div 2 \times 10^7}$ GeV.
 This limit is by a factor of 2.8 more stringent than a limit obtained
 with a previous analysis.
  \end{abstract}

 
\section{Introduction}
High-energy neutrinos are likely produced in violent
processes in the Universe. Many theoretical models predict
that neutrinos are generated by hadronic processes within
high energy astrophysical sources such as active galactic
nuclei (AGN), supernova remnants or gamma ray bursts.
Individual sources might be too weak to produce an unambiguous
directional signal, however the total neutrino flux from all
sources could produce a detectable diffuse neutrino signal.
Astrophysical neutrinos generated in top-down models are, by
definition, of diffuse nature. To date the highest sensitivities
to diffuse neutrino fluxes in a range 10 TeV $\div$ 100 PeV are achieved 
 with the NT200(Baikal) \cite{APP06} and AMANDA \cite{AMANDAHE1,AMANDAHE2} neutrino telescopes.

The Baikal Neutrino Telescope NT200 is operating in Lake Baikal at
a depth of 1.1 km and is taking data since 1998. 
Since 2005, the upgraded 10-Mton scale
detector NT200+ is in operation. Detector configuration and
performance have been described elsewhere 
\cite{APP1,RW,RICAP07,VLVNT08_st}.
Due to high water transparency and low light scattering, the detection volume
of NT200 for high energy $\nu_e$, $\nu_{\mu}$ and $\nu_{\tau}$ events significantly
exceeds the instrumented volume. Our previous analysis \cite{APP06} of 1038 live-days data, 
collected in the years 1998-2002  with NT200, has allowed to set the limits on diffuse neutrino fluxes
predicted by several theoretical models. Here we discuss a new analysis method which is based on
an energy and space-angular reconstruction of high-energy cascades and present limits
on diffuse neutrino fluxes  which are improved by a factor of about three over the previous ones.

  \begin{table*}[th]
  \caption{Expected number of events $N_{model}$, detection energy range which contains
the central 90\% of the expected signal $\Delta E_{90\%}$, median energy of the expected signal $\bar{E_{\nu}}$,   
and model rejection factors $\eta=n_{90\%}/N_{model}$ for models of astrophysical neutrino sources.}
  \label{table_wide}
  \centering
  \begin{tabular}{|@{}c|c|c|c|c|c|}
  \hline
 & \multicolumn{4}{c|}{BAIKAL}  & AMANDA \\
\hline
Model & $N_{model}(\nu_e+\nu_{\mu}+\nu_{\tau})$ & $\Delta E_{90\%}$ & $\bar{E_{\nu}}$ & $n_{90\%}/N_{model}$ & 
$n_{90\%}/N_{model}$  \\
\hline
  S05 & 0.7 & 100 TeV $\div$ 30 PeV & 2 PeV & 3.4 & 1.6  \\
  P $p\gamma$ & 4.4 & 320 TeV $\div$ 160 PeV & 6 PeV & 0.5 & 0.3  \\
  M $pp+p\gamma$ & 1.7 & 20 TeV $\div$ 500 PeV & 15 PeV & 1.4 & 1.2  \\
  MPR & 1.4 & 160 TeV $\div$ 100 PeV & 3 PeV & 1.8 & 0.9  \\
  SeSi & 2.4 & 1 PeV $\div$ 50 PeV & 10 PeV & 1.0 & -  \\
  \hline
  \end{tabular}
  \end{table*}

\section{The analysis method}
The BAIKAL survey for high energy neutrinos searches
for bright cascades produced at the neutrino interaction
vertex in a large volume around and below the telescope.
The main background source are atmospheric muons, with
a flux $10^6$ times higher than that of atmospheric neutrinos.
We select events with high multiplicity of hit channels $N_{\mbox{\small hit}}$,
corresponding to bright cascades. 
To separate high-energy neutrino events
from background events, a cut
to select events with upward moving light signals has been developed.
We define for each event
$t_{\mbox{\footnotesize min}}=\mbox{min}(t_i-t_j)$,
where $t_i, \, t_j$ are the arrival times at channels $i,j$ 
on each string, and the minimum over all strings is calculated.
Positive and negative values of $t_{\mbox{\footnotesize min}}$ correspond to 
upward and downward propagation of light, respectively. 
We require
\begin{equation}
t_{\mbox{\footnotesize min}} > -10 \mbox{ns}.
\end{equation}
This cut accepts only time patterns corresponding to up-ward traveling
light signals. It rejects most events from brem-cascades produced
by downward going muons since the majority of muons is close to the
vertical; they would cross the detector or pass nearby and generate a
downward time pattern. Only few muons with large zenith angles
may escape this cut and illuminate the array by their own Cherenkov
radiation or that from bright cascades from below. 

The energy spectrum of neutrinos from galactic and cosmological
sources or from the decay of topological defects is expected
to have a significantly flatter shape than the spectrum of
atmospheric muons and neutrinos. 
This gives rise to different $N_{hit}$ and cascades energy distributions.
Results of a search for high-energy neutrinos,
based on $N_{hit}$ as a rough indicator of the energy deposited 
in the effective detection volume 
were published in \cite{APP06}.
Here we present results of an extended analysis which is based 
on a full reconstruction of cascades parameters.

\subsection{Cascade parameters reconstruction}
We use a two step procedure for cascade parameters reconstruction
which is applied to events with $\ge 5$ hit channels.
At the first step, using the time information of hit channels,
the coordinates of the cascade vertex are reconstructed
by minimizing
\begin{equation}
\chi_t^{2}=\frac{1}{(N_{hit}-4)}\sum_{i=1}^{N_{hit}}\frac{(T_i(x,y,z,t_0)-t_i)^2}{\sigma_{ti}^2}.
\end{equation}
Here $t_i$ is the  time measured by the $i$-th channel, $T_i$ is the expected arrival time of Cherenkov photons
induced by a cascade with $\vec{r}_{sh}(x,y,z)$ coordinates, and 
$\sigma_{ti}$ are the timing errors.
At the next step, taking into account the found coordinates $\vec{r}_{sh}(x,y,z)$,  
we reconstruct the cascades energy $E_{sh}$, as well as zenith and azimuth angles   
$\theta$ and $\varphi$ of cascade axis by maximizing the likelihood function
\begin{equation}
 L_A=\prod_{i=1}^{N_{hit}} p_i(A_i,E_{sh},\vec{\Omega}_{sh}(\theta,\varphi)).
\end{equation}
Here $p_i(A_i,E_{sh},\vec{\Omega}_{sh}(\theta,\varphi))$ is the probability
to detect a signal of amplitude $A_i$ by the $i$-th hit channel. 
These probability functions are calculated with taking into account the absorption
and scattering of light in water and the relative orientations of optical modules and cascades.

 \begin{figure*}[!t]
\includegraphics[width=2.5in]{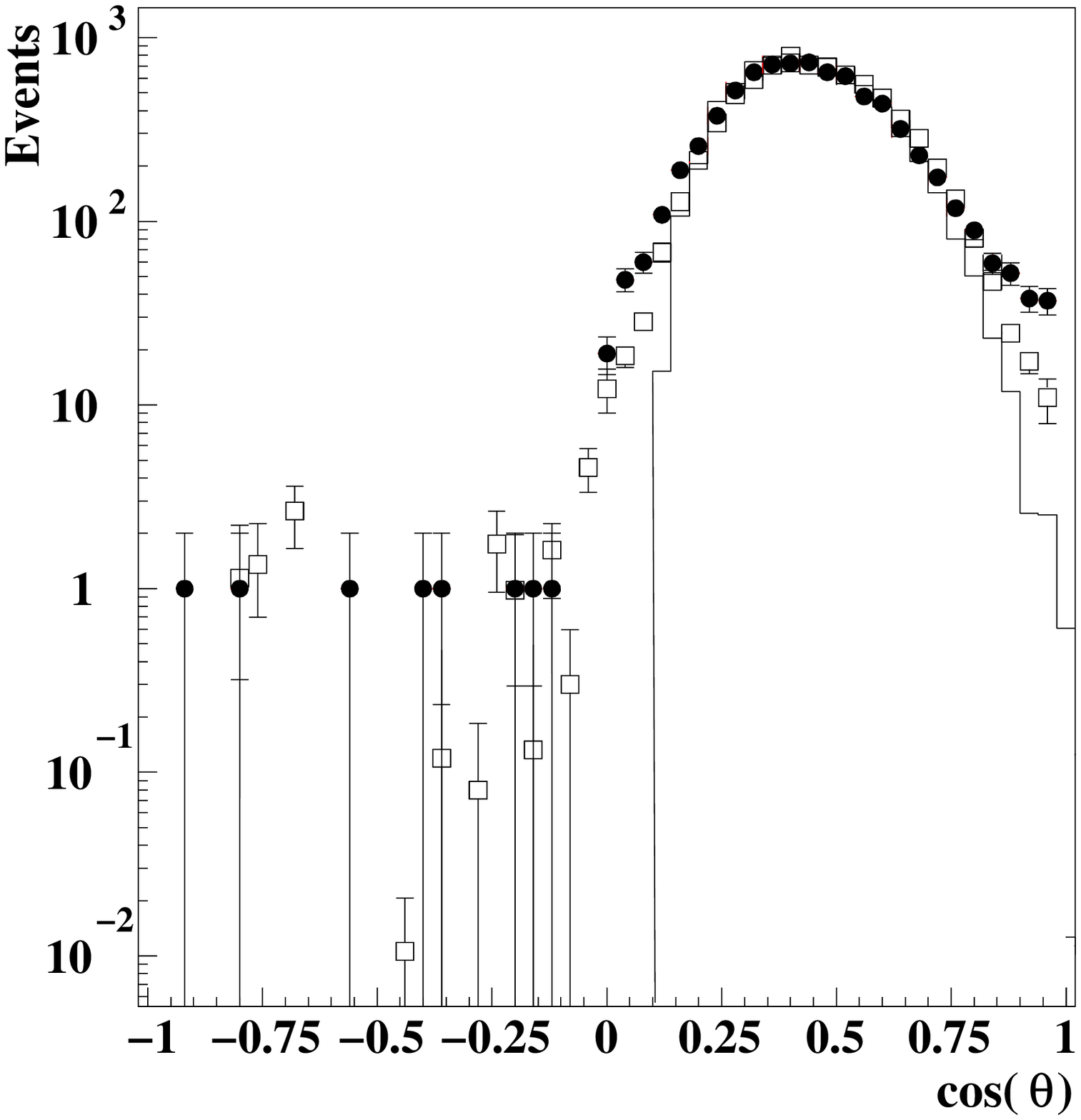}
              \hfil
\includegraphics[width=2.5in]{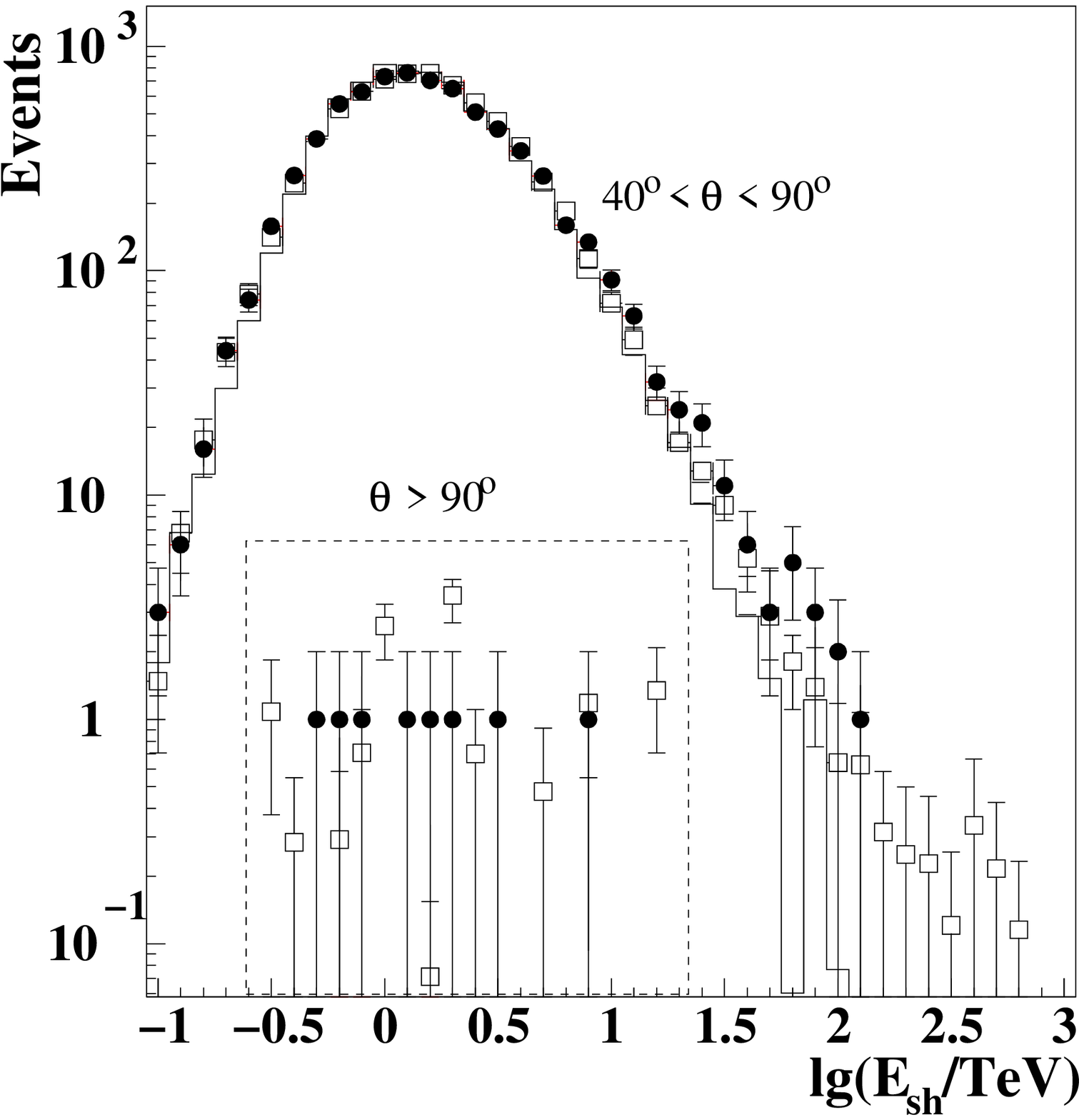}
\\   
   \caption{Left: Reconstructed cascade zenith-angule distribution for data (dots) and for MC-generated
atmospheric muons (boxes); true MC zenith-angule distribution is given as histogram.
Right: Reconstructed cascade energy distribution for data (dots) and for MC-generated
atmospheric muons (boxes); true MC energy distribution is given as histogram.
}
   \label{double_fig}
 \end{figure*}

\subsection{High-energy neutrino simulation}
The number of expected events during observation time $T$ is
$$
N_{\nu}=T \int d\vec{\Omega}\int dE_{sh}V_{eff}
(\vec{\Omega},E_{sh})\sum_k \int
 N_{A} \rho_{H_2O}
$$
\begin{equation}
\label{h2}
 \times \frac{d\sigma_{\nu k}}{dE_{sh}}
\Phi_{\nu}(\vec{\Omega},E_{\nu},X)dE_{\nu},
\end{equation}
$$
X(\vec{\Omega})=\int_{0}^{L} \rho_{earth}(l)dl, 
$$
where $\Phi_{\nu}(\vec{\Omega},E_{\nu},X)$ is the flux of high-energy
neutrinos with energy $E_{\nu}$ in the vicinity of the detector,
$\vec{\Omega}$ -- the neutrino direction, $X(\vec{\Omega})$ -- the
thickness of matter encountered by the neutrino on its passage
through the Earth, $E_{sh}$ -- the energy of secondary cascades,
$V_{eff}(\vec{\Omega},E_{sh})$ -- the detection volume. The index
$\nu_i$ indicates the neutrino type and $k=$1,2 corresponds to CC- and NC-interactions,
respectively. $N_A$ is the Avogadro number and $\rho_{H_2O}$ the water density.

A MC-code is used to solve Eq. (\ref{h2}), with the boundary conditions for neutrino fluxes
$\Phi_{\nu_i}(E,0)=A_{\nu_i}f_{\nu_i}(E)$, where $f_{\nu_i}(E)$ is a diffuse
AGN-like flux or other predicted UHE neutrino fluxes, and a $A_{\nu_i}$ a
normalization coefficient. For neutrino interactions we used cross-sections from \cite{Reno96}.
The neutrinos are propagated through the Earth assuming the density profile of the
Preliminary Reference Earth Model \cite{EARTH}. Although a flavor ratio of 
$\nu_e$:$\nu_{\mu}$:$\nu_{\tau} \approx $1:2:0 is predicted for generic neutrino fluxes
at astrophysical sources, equal fractions of all three neutrino flavors are expected
at Earth because of neutrino oscillations. Throughout this paper we assumed a neutrino
flavor ratio at Earth of $\nu_e$:$\nu_{\mu}$:$\nu_{\tau}=$1:1:1 and the same shape
of energy spectra $f_{\nu}(E)$ for all neutrino flavors,
as well as a flux ratio for neutrino and antineutrino of 
$\nu/\bar{\nu}=$1\footnote{A violation of this assumption (e.g. for neutrino 
production in $p\gamma$ interactions) has a small influence on the result due 
to the similarity of $\nu$ and $\bar{\nu}$ cross-sections in our energy 
range.}.

The detector response to Cherenkov radiation of high energy
cascades was simulated taking into account the effects
of absorption and scattering of light, 
as well as light velocity dispersion in water.
We also implemented the longitudinal development of cascades.
For electron cascades with $E_{\mbox{\footnotesize sh}}>$2$\times$10$^7$ GeV
and for hadronic cascades with $E_{\mbox{\footnotesize sh}}>$10$^9$ GeV, 
the increase in cascade length due to the LPM effect \cite{LPM1}
was approximated as $E^{1/3}$ according to \cite{LPM}.

\subsection{Atmospheric muon simulation}
Downward going atmospheric muons 
are the most important source of background. 
The simulation chain of these muons starts with 
cosmic ray air shower generation using the CORSIKA
program \cite{CORSIKA} with the QGSJET \cite{QGSJET}
interaction model and
the primary composition and spectral slopes
for individual elements taken from \cite{Smooth}.
Atmospheric muons are propagated through the water
using the MUM program \cite{MUM}. 
During passage through the detection volume the detector response 
to Cherenkov light from all muon energy loss processes
is simulated.
A total of $1.2 \times 10^9$ background events, equivalent to
3671 live days, has been simulated, with standard optical 
parameters of Baikal water. 
The cascade reconstruction procedure was applied to
simulated background events with $N_{hit}>$15 which obey the
condition (1). To reject events induced by muon bundles  the following additional
cuts were applied: $N_{hit}>$18, $\chi_t^{2}<$3 and $L_A<20$.
Zenith-angule and energy distributions of the selected events are shown in Fig.1 
and are discussed in the next section.

\section{Results}
Within the 1038 days of the detector live time,
$3.45 \times 10^8$ events with $N_{\mbox{\small hit}} \ge 4$ have been 
recorded.
For this analysis we used 18384 events with hit channel multiplicity
$N_{\mbox{\small hit}}>$15 and $t_{\mbox{\footnotesize min}}>$-10 ns.
As it was shown in \cite{APP06} the data are consistent with simulated background
for both $t_{\mbox{\footnotesize min}}$ and $N_{\mbox{\small hit}}$ 
distributions. 

A full cascade reconstruction algorithm 
(for vertex, direction, energy) was applied to
the data \cite{AVR}. Cuts were then placed on this reconstructed
cascade energy to select neutrino events.

The reconstructed zenith-angule and energy distributions of data  
are shown in Fig.1 (dots). Eight events were reconstructed as upward going
cascades (zenith angle $\theta > 90^{\circ}$ in the left panel and the distribution in the dashed box in right panel).
Also the MC-generated (histograms) and reconstructed (boxes) zenith-angule and 
energy distributions from simulated atmospheric muons are shown in
Fig.1; 12 upward reconstructed cascade-like events are expected. 
As seen from Fig.1, within systematic and statistical
uncertainties there is no significant excess above the background from atmospheric muons. 
We introduce the following final neutrino signal cuts on the cascade energy: E$_{sh}>$130 TeV and  E$_{sh}>$10 TeV
for downward (40$^{\circ}<\theta<$90$^{\circ}$) and upward going cascades, respectively (see for details \cite{AVR}).
Furthermore, events which fulfil selection requirements used in our previous analysis \cite{APP06}
are also considered as neutrino candidates.
With zero observed events and 2.3$\pm$1.2 expected background
events, a 90\% confidence level upper limit on the number
of signal events of $n_{90\%}=$2.4 is obtained.

A model of astrophysical neutrino sources, for which the total number
of expected events, $N_{model}$, is larger than 
$n_{90\%}$, is ruled out at 90\% CL. 
Table I represents event rates,
detection energy range which contains
the central 90\% of the expected signal, median energy of the expected signal,
 and model rejection factors (MRF) 
$n_{90\%}/N_{model}$ for models of astrophysical neutrino sources 
obtained from our search, as well as model rejection factors obtained recently
by the AMANDA collaboration \cite{AMANDAHE1,AMANDAHE2}.
The model by Stecker \cite{S05} labeled ``S05'',
represents models for neutrino production in the central
region of Active Galactic Nuclei. Further shown in Table I
are models for neutrino production in AGN jets:
calculations by Protheroe \cite{P} and by Mannheim \cite{M},
which include neutrino production through $pp$ and $p\gamma$ collisions
(models ``P $p\gamma$'' and ``M $pp+p\gamma$'', respectively),
as well as an evaluation of the maximum flux due to a superposition of
possible extragalactic sources by Mannheim, Protheroe and Rachen
\cite{MPR} (model ``MPR'') and a prediction for the diffuse
flux from blazars by Semikoz and Sigl \cite{SeSi} ``SeSi''.
As can be seen from Table I the model ``P $p\gamma$'' is ruled out with
$n_{90\%}/N_{model} =$ 0.5. 

For an $E^{-2}$ behaviour of the neutrino spectrum and a flavor ratio 
$\nu_e:\nu_{\mu}:\nu_{\tau}=1:1:1$, the 90\% C.L. upper limit on the all-flavor 
neutrino flux obtained with the Baikal neutrino telescope  
NT200 is: 
\begin{equation}
E^2\Phi<2.9 \times 10^{-7} \mbox{cm}^{-2}\mbox{s}^{-1}\mbox{sr}^{-1}\mbox{GeV},
\end{equation}
for $20\,$TeV$\,<\,$E$_{\nu}\,<20\,$PeV.
Fig.2 shows our upper limit on 
the all-flavor $E^{-2}$ diffuse flux, which is a significant
improvement of the earlier obtained limit \cite{APP06}.
Also shown are the limits obtained by AMANDA 
\cite{AMANDAHE1,AMANDAHE2}
and Pierre Auger Observatory \cite{Auger}, theoretical bounds obtained by 
Berezinsky \cite{Ber4}, 
by Waxman and Bahcall \cite{WB1}, by Mannheim et al.(MPR) 
\cite{MPR}, as well as the atmospheric conventional 
neutrino \mbox{fluxes \cite{VOL}}. 

 \begin{figure}[t]
\includegraphics[width=3.0in]{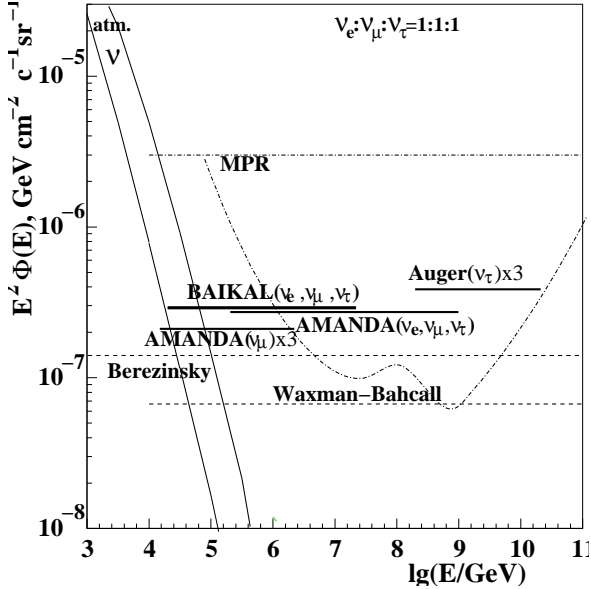}

   \caption{All-flavor neutrino flux limits and theoretical bounds (see text).
}
   \label{singl_fig}
 \end{figure}

\section{Conclusion}
The neutrino telescope NT200 in Lake Baikal is taking data
since April 1998. Due to high water transparency and low
light scattering, the detection volume of NT200 for high energy
neutrino events is several Megatons and significantly exceeds the 
instrumented volume. This results in a high sensitivity to diffuse
neutrino fluxes from extraterrestrial sources -- more than an
order of magnitude better than that of underground searches and
similar to the published limits of the AMANDA neutrino telescope.
The upper limit obtained for a diffuse all-flavor neutrino flux
with $E^{-2}$ shape is $E^2 \Phi = 2.9 \times 10^{-7}$\,cm$^{-2}$\,s$^{-1}$\,sr$^{-1}$\,GeV. 


\section{Acknowledgments}
{\small
This work was supported in part by the Russian Ministry of Education
and Science, by the German Ministry of Education and Research,
by the Russian Found for Basic Research (grants 08-02-00432-a,
07-02-00791, 08-02-00198, 09-02-10001-k, 09-02-00623-a), by the grant of
the President of Russia NSh-321.2008-2 and
by the program "Development of Scientific Potential in Higher Schools"
(projects 2.2.1.1/1483, 2.1.1/1539, 2.2.1.1/5901).
}



\begin{thebibliography}{99}
\bibitem{APP06} V.~Aynutdinov et al., {\it Astropart. Phys., vol. 25, pp. 140-150, 2006}.
\bibitem{AMANDAHE1} A.~Achterberg et al., {\it Phys. Rev., vol. D76, p. 042008, 2007}.
\bibitem{AMANDAHE2} M.~Ackermann et al., {\it Astrophys. J., vol. 675, p. 1014, 2008}.
\bibitem{APP1} I.~Belolaptikov et al., {\it Astropart. Phys., vol. 7, p. 263, 1997}.
\bibitem{RW} V.~Aynutdinov et al., {\it Nucl. Instrum. Methods, vol. A567, p. 433, 2006}.
\bibitem{RICAP07} V.~Aynutdinov et al., {\it Nucl. Instrum. Methods, vol. A588, p. 99, 2008}.
\bibitem{VLVNT08_st} V.~Aynutdinov et al., {\it Nucl. Instrum. Methods, vol. A602, p. 14, 2009}.
\bibitem{Reno96} R.~Gandhi et al., {\it Astropart. Phys., vol. 5, p. 81, 1996};
R. Gandhi et al., {\it Phys. Rev., vol. D58, p. 093009, 1998}.
\bibitem{EARTH} A.M.~Dziewonski, D.L.~Anderson, {\it Phys. Earth Planet. 
Interiors, vol. 25 p. 297, 1981}.
\bibitem{LPM1} A.~Migdal, {\it Phys. Rev., vol. 103, no. 6, p. 1811, 1956}.
\bibitem{LPM} J.~Alvarez-Muniz, E.~Zas, {\it Phys. Lett., vol. B411, p. 218, 1997}.
\bibitem{CORSIKA} J.~Capdevielle et. al., {\it KfK Report 4998, 
Kernforschungszentrum, Karlsruhe, 1992}.
\bibitem{QGSJET} N.N.~Kalmykov, S.S.~Ostapchenko and A.I.~Pavlov,
{\it Nucl. Phys. (Proc. Suppl.), vol. B52, p. 17, 1997}.
\bibitem{Smooth} B.~Wiebel-Smooth, P.~Biermann and Landolt-Bornstein, 
{\it Cosmic Rays, vol. 6/3c, pp. 37-90, Springer Verlag, 1999}.
\bibitem{MUM} E.V.~Bugaev et al., {\it Phys. Rev., vol. D64, p. 074015, 2001}.
\bibitem{AVR} A.~Avrorin et al., {\it Astronomy Letters}, in press.
\bibitem{S05} F.W.~Stecker, {\it Phys. Rev., vol. D72, p. 107301, 2005}.
\bibitem{P} R.J.~Protheroe, {\it arXiv.org:astro-ph /9612213}.
\bibitem{M} K.~Mannheim, {\it Astropart. Phys., vol. 3, p. 295, 1995}.
\bibitem{MPR} K.~Mannheim, R.~Protheroe and J.~Rachen, 
{\it Phys. Rev., vol. D63 p. 023003, 2001}.
\bibitem{SeSi} D.~Semikoz and G.~Sigl, {\it arXiv.org:hep-ph/0309328}.
\bibitem{Auger} J.~Abraham et al., {\it Phys. Rev. Lett., vol. 100, p. 211101, 2008}.
\bibitem{Ber4} V.~Berezinsky, {\it arXiv.org: astro-ph/0505220}.
\bibitem{WB1} E.~Waxman and J.~Bahcall, {\it Phys. Rev., vol. D59, p. 023002, 1999}.
\bibitem{VOL} L.~Volkova, {\it Yad. Fiz., vol. 31, p. 1510, 1980}.

  \end{thebibliography}
\end{document}